# **Ultradense Deuterium**

F. Winterberg

University of Nevada, Reno

**Abstract**

An attempt is made to explain the recently reported occurrence of ultradense deuterium as an isothermal transition of Rydberg matter into a high density phase by quantum mechanical exchange forces. It is conjectured that the transition is made possible by the formation of vortices in a Cooper pair electron fluid, separating the electrons from the deuterons, with the deuterons undergoing Bose-Einstein condensation in the core of the vortices. If such a state of deuterium should exist at the reported density of about 130,000 g/cm$^3$, it would greatly facility the ignition of a thermonuclear detonation wave in pure deuterium, by placing the deuterium in a thin disc, to be ignited by a pulsed ultrafast laser or particle beam of modest energy.

**1. Introduction**

Rydberg matter is made up from highly excited atoms, where the outermost electron makes a circular motion around its atomic core. The existence of Rydberg matter was in 1980 first predicted by E.A. Manykin, Ozhovan and Puluektov [1]. But it was a research group in Sweden at University of Gothenburg, under the leadership of Leif Holmlid, which has recently announced it had discovered an ultradense form of deuterium by a phase transition from a Rydberg matter state of deuterium, a million times more dense than liquid deuterium [2]. Because this claim is so extraordinary, it must be taken with a great deal of skepticism. But since Leif Holmlid has an established record of publications about Rydberg matter in the refereed scientific literature the claim cannot be easily dismissed. It is the purpose of this communication to explore the question if such an unusual state of matter might exist.

**2. Towards a possible explanation**

The publication [2] by the Swedish research group contains a few hints for a possible explanation of ultradense deuterium:

1. The isothermal constant pressure transition to the ultradense phase is possible only for deuterium (D), not for hydrogen (H).
2. The two phases, the normal and ultradense, co-exist in a mixture of both.
3. The superdense phase is formed in pores with $Fe_2O_3$ acting as a catalyst [3].
4. The superdense phase can be reached only through a transition from Rydberg deuterium matter.
5. In the ultradense phase the kinetic energy and angular momentum of the electrons are "switched" to the energy and angular momentum of the deuterons.

To 1: This points to a configuration where the deuterons form a fluid undergoing Bose-Einstein condensation, not possible for protons.

To 2: Both phases, the normal and the superdense, can be described by a van der Waals equation of state.

To 3: Because $Fe_2O_3$ is a ferromagnet, this suggests the ordering effect of a magnetic field is needed to reach the ultradense phase.

To 4: Because Rydberg matter can be described as composed of discs, with the outer electron moving in a circle around to atomic core, in our case around the deuterium nucleus, this suggests that the ultradense state is a tower of piled up of deuterium Rydberg atoms.

To 5: Since under normal conditions it is not possible to have a configuration where the deuterons move around the electrons, this points into the direction of a large effective mass for the electrons, possible if the electron fluid forms vortices, because vortices have a large effective mass.

### 3. Vortex model

A detailed theoretical study of possible metallic superfluid hydrogen (H) and deuterium (D) has been made by Ashcroft [4], but with no definite conclusion and with no prediction of an ultrahigh phase of deuterium. Because of the very complex nature of the posed problem a different approach for its solution is here suggested: Using the information provided by experiment, one may try to guess the correct configuration. To get an idea how the ultradense configuration may look, we display in **Fig. 1** a hexagonal cluster of 19 planar Rydberg atoms with their circular electron orbits, and in **Fig. 2** the distribution of the valance electrons in this cluster [5,6]. **Fig. 3** finally shows how the clusters can be piled up into a tower.

With this information we guess the shape of the ultradense configuration shown in **Fig. 4**. It consists of a vortex lattice, with the vortices made up from electron Cooper pairs, and with a string of superfluid deuterons centered in the core of these vortices. Because the deuterons obey Bose-Einstein condensation, the axial distances between the deuterons in each string can become small. The exchange forces of the valence electrons act radially inward and compress the vortex lattice radially. To avoid a space charge build-up in the axial direction, the deuteron strings are subject of an axial stress, bringing them closer together until they are separated by roughly the same distance as the distance in between the electrons.

The large effective mass of the vortices follows from their equation of motion. For their equation of motion, the Newtonian law of motion – mass × acceleration = force -, is replaced by mass* × velocity = force*, where m* and f* are an effective mass and force [7]. If compared with Newton's equation, the vortex equation of motion acts as if the vortices have an infinite effective mass, which in Newton's mechanics leads to a constant velocity under the influence of a force of finite strength. The same is true in vortex dynamics for the motion of vortex under a finite (effective) force.

The occurrence of vortices under the described condition can best be seen from the hydrodynamic formulation of quantum mechanics, obtained by the Madelung transformation of the Schrödinger equation [8]:

$$\frac{\partial \mathbf{v}}{\partial t} + \frac{1}{2}\text{grad}\mathbf{v}^2 - \mathbf{v} \times \text{curl}\mathbf{v} = -\frac{1}{m}\text{grad}[V + Q] \qquad (1)$$

where **v** is the velocity of the quantum fluid, further V the electric and Q the quantum potential. For distances large compared to the de Broglie length ℏ/mv, Q can be neglected. There then with curl**v** = 0, one obtains from (1) the potential vortex solution

$$v = v_0 \left(\frac{r_0}{r}\right), \quad r > r_0 \qquad (2)$$

where for > ℏ/mv, the quantum potential can be neglected.

## 4. Dynamical considerations

If the kinetic energy and angular momentum of electron is "switched" to the energy and angular momentum of the deuteron, one must have

$$(1/2)mv_B^2 = (1/2)MV_d^2 \qquad (3)$$

$$mr_B v_B = Mr_d V_d \qquad (4)$$

where $r_B$ is the Bohr radius, $v_B=\alpha c$ ($\alpha=1/137$) the electron velocity at $r = r_B$, m and M the electron and deuteron mass, $r_d$ the radius at which the deuteron moves on a circular orbit with the velocity $V_d$. From (3) and (4) one obtains (setting $M = 2M_H$, $M_H$ proton mass):

$$r_d = r_B (m/M)^{1/2} = (r_B/\sqrt{2})(m/M_H)^{1/2} = 1.65\times10^{-2}\, r_B = 8.25\times10^{-10}\, cm \qquad (5)$$

$$V_d = v_B (m/M)^{1/2} = (v_B/\sqrt{2})(m/M_H)^{1/2} = 1.2\times10^{-4}\, c = 3.6\times10^6\, cm/s \qquad (6)$$

Setting $r_0 = r_d$ where $r_0$ is the vortex core radius on obtains at this radius for the electron velocity

$$v = v_0 = (M/m)V_d = (M/m)^{1/2}\, v_B = \sqrt{2}\, v_B (M_H/m)^{1/2} = 0.43c \qquad (7)$$

or

$$v_0/v_B = (M/m)^{1/2} = 60.5 \qquad (8)$$

If the electron fluid is made up of Cooper pairs, these relations are going to change, but not in a significant way, with $r_d$ changed by $\sqrt{2}$ to $r_d = r_B (m/M_H)^{1/2} = 2.34\times10^{-2}\, r_B$, and $V_d$ changed to $V_d = v_B (m/M_H)^{1/2} = 5.1\times10^6\, cm/s$, and $v_0/v_B$ changed to $v_0/v_B = 43$.

## 5. Linear Atom

As shown in **Fig. 4** the deuterons together with electrons form a tower. The height of this tower, which may be called a linear atom shall be equal to *h*, and the radius of this cylindrical linear atom equal to r. This linear atom has the electric potential

$$V = 2q\ln\left(\frac{h}{r}\right) \tag{9}$$

where q is the charge per unit length of the linear atom. If the deuterons are equally distributed along the tower by the distance r, one can write for (9)

$$V \simeq \frac{2e}{r}\ln\left(\frac{h}{r}\right) \tag{10}$$

The radially directed electric field is then

$$E = -\frac{\partial V}{\partial r} = \frac{2e}{r^2}\left[1 + \ln\left(\frac{h}{r}\right)\right] \tag{11}$$

We compute the smallest electron radius as a linear Bohr atom, equating the centrifugal force with the electric force

$$\frac{mv^2}{r} = \frac{2e^2}{r^2}\left[1 + \ln\left(\frac{h}{r}\right)\right] \tag{12}$$

to be supplemented by

$$mrv = \hbar \tag{13}$$

Form (12) and (13) we have

$$r = \frac{r_B}{2\left[1 + \ln\left(\frac{h}{r}\right)\right]} \tag{14}$$

Since ln (h/r) does not change very much with r, we can equate there r with r = $r_d$, and obtain

$$\ln\left(\frac{h}{r_d}\right) \simeq \frac{1}{\sqrt{2}}\left(\frac{M_H}{m}\right)^{1/2} - 1 \simeq 29 \tag{15}$$

hence

$$\frac{h}{r_d} \simeq 4 \times 10^{12} \tag{16}$$

or $h \simeq 3.3 \times 10^3$ cm, a completely unrealistic value.

The situation is very much changed if the electrons form Cooper pairs. There one first has to set in (12) m→2m but also e→2e, leaving (12) unchanged. Setting e→2e is needed to compensate the charge of the Cooper pairs with the charge of two deuterons. But (13) is changed into

$$2mrv = \hbar \tag{17}$$

whereby (14) is changed into

$$r = \frac{r_B}{8\left[1 + \ln\left(\frac{h}{r}\right)\right]} \tag{18}$$

Then, equating r with $r_d = r_B (m/M_H)^{1/2}$, one finds that

$$\ln\left(\frac{h}{r_d}\right) \simeq \frac{1}{8}\left(\frac{M_H}{m}\right)^{1/2} - 1 \tag{19}$$

or

$$h \simeq 70\ r_d \simeq 1.65\ r_B = 8 \times 10^{-9}\ \text{cm}$$

Here the tower is made up of 2×70 = 140 deuterons.

### 6. Exchange forces

The assumption that the deuterons can lump into pairs to compensate the double charge of the Cooper pairs needs a force. It is different from the force by the Cooper pairs which results from the electron-phonon interaction. The electron-phonon interaction requires the deuteron vortex lattice, leading to the occurrence of the phonons. To lump the deuterons as suggested, must for this reason be explained by another strong force. It is suggested that this force is the quantum mechanical exchange force between two deuterons. If the electron wave function at the position $\mathbf{r}_1$ of one deuteron is given by $\psi_a(\mathbf{r}_1)$ and by a second electron at the position $\mathbf{r}_2$ of another deuteron by $\psi_b(\mathbf{r}_2)$, this leads to an exchange force to result from the integral

$$J_{ab} = \iint \psi_a^*(\mathbf{r}_1)\psi_b^*(\mathbf{r}_2) \frac{e^2}{|\mathbf{r}_1 - \mathbf{r}_2|} \psi_b(\mathbf{r}_1)\psi_a(\mathbf{r}_2) d\mathbf{r}_1 d\mathbf{r}_2 \qquad (20)$$

As in the Heitler-London theory for the chemical bonding, it results in a lowering of the energy. This exchange energy must be summed up between all deuteron pairs, but will be strongest between neighboring pairs. One obtains:

$$H_{exch} = -2 \sum_{i,k} J(|\mathbf{R}_j - \mathbf{R}_k|) \qquad (21)$$

where

$$J(|\mathbf{R}_j - \mathbf{R}_k|) = \iint \psi_j^*(|\mathbf{R}_j - \mathbf{r}_j|) \psi_k^*(|\mathbf{R}_k - \mathbf{r}_k|)$$

$$\frac{e^2}{(|\mathbf{r}_j - \mathbf{r}_k|)} \psi_k(|\mathbf{R}_k - \mathbf{r}_j|)\psi_j(|\mathbf{R}_j - \mathbf{r}_k|) d\mathbf{r}_j d\mathbf{r}_k \qquad (22)$$

and where the interacting pair of deuterons are positioned at $\mathbf{R}_j$ and $\mathbf{R}_k$. This exchange energy between two layers of the linear atom is negative and is calculated to be:

$$H_{exch} \sim -1.15\ e^2/r_d \qquad (23)$$

### 7. Stability

Even though the existence of ultradense deuterium cannot be excluded for theoretical reasons, the stability of the conjectured configuration in **Fig. 4** is far from certain. In **Fig. 4** it is assumed that the vortices have all the same circulation, both by magnitude and direction. In this configuration the electron currents cancel each other and the vortex lattice behaves like a superconductor. But there is an alternative configuration, where the circulation of adjacent vortices is equal but opposite. There the currents attract neighboring vortices, but according to the laws of vortex dynamics [7] the vortices undergo microscopic motions which could make this configuration unstable, even though it is magnetically favorable.

## 8. Thermonuclear ignition

The idea to use the ultradense configuration of deuterium for the thermonuclear ignition of deuterium was not lost on the Swedish group who reported the existence of ultradense deuterium [2, 9, 10, 11]. Depending on its stability it would without any doubt to be ideal for inertial confinement fusion. With a thousandfold compression, an energy upto 10 MJ is needed for ignition of the deuterium-tritium reaction. The ignition energy goes in direct proportion of the ignition temperature and in inverse proportion of the target density, with the ignition temperature of deuterium about 10 times larger. Together with a further $10^3$ fold increase in the target density, this would bring down the ignition energy by a factor of $10^5$, from $10^8$ J to about 1kJ. With a bremsstrahlung radiation loss time at $10^6$ times solid density of deuterium at a temperature of ~ $10^8$ K about $2\times10^{-13}$ sec, the ignition power required would be less than 10 petawatt, still within what may be technically feasible, both with lasers and charged particle beams.

The presumably high target density suggests a pancake type target, made from a thin layer of ultradense deuterium. Because of the $\rho r > 10$ g/cm$^2$ condition for the detonation of pure deuterium, the thickness of the ultradense layer of deuterium would have to be $\geq 10^{-4}$ cm, condensed into the ultradense state from a sheet 100 times thicker layer. A possible target configuration is shown in **Fig. 5**. It has the shape of a flat lens, containing liquid deuterium. During its transformation in the ultradense phase, the spatial dimension of the lens is decreased 100 fold, with the latent heat removed by heat conduction. The ignition can be initiated in the central spot of the lens, either by a petawatt laser or a charged particle beam. With the energy requirement for ignition decreased by a factor $10^5$, the size of the capacitor of the super-Marx generator proposed for the ignition of a pure deuterium reaction would be smaller by the same factor, reducing its linear dimension $10^{5/3} \simeq 50$ fold, from one kilometer in length to 20 meter, and from a diameter of 20 meters down to less than 1 meter.

The condensation into the superdense state follows a constant pressure isotherm of a van der Waals equation of state. According to the first law of thermodynamics the removal of heat is equal to

$$w = \int_{\rho_0}^{\rho_1} p\, d\left(\frac{1}{\rho}\right) \qquad (24)$$

According to the second law of thermodynamics the integral (24) must be evaluated by integrating it along a reversible path. This means we have to set $p/\rho = (R/A)T$ (R universal gas constant, A=2 for deuterium). Hence

$$w = \frac{RT}{A} \int_{\rho_0}^{\rho_1} \frac{d\rho}{\rho} = \frac{RT}{2} \ln\left(\frac{\rho_1}{\rho_0}\right) \qquad (25)$$

For $\rho_1 = 10^6 \rho_0$, this is

$$W = 3RT \ln 10 \; [\text{erg/g}] \qquad (26)$$

at low temperatures a comparatively small amount of heat. Since the deuterium is in a superconducting – superfluid state, this heat can be easily removed.

**Conclusion**

If as reported the state of ultradense deuterium exists, and if it is sufficiently stable to exist long enough, it could become for the release of nuclear energy as important as was the discovery of nuclear fission by Hahn and Strassmann. It is the purpose of this note that on purely theoretical grounds an ultradense state of deuterium cannot be easily dismissed.

**Figures**

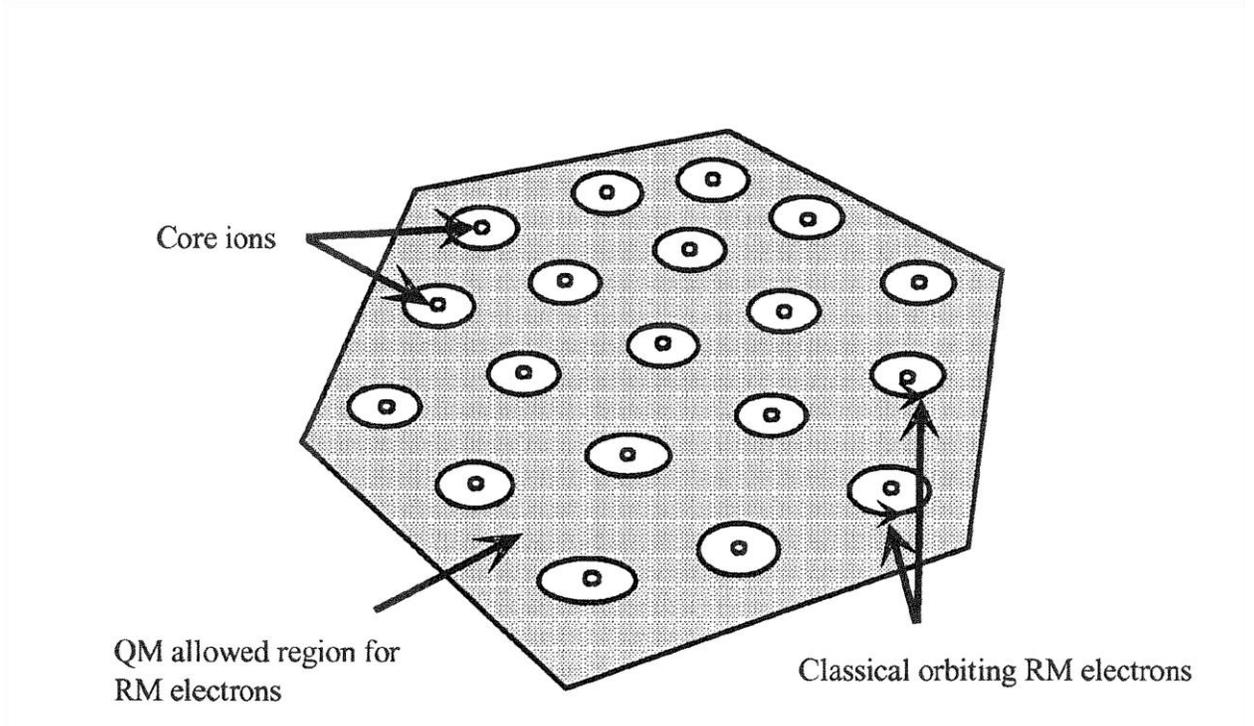

**Fig. 1**. Rydberg matter cluster of 19 Deuterons [6].

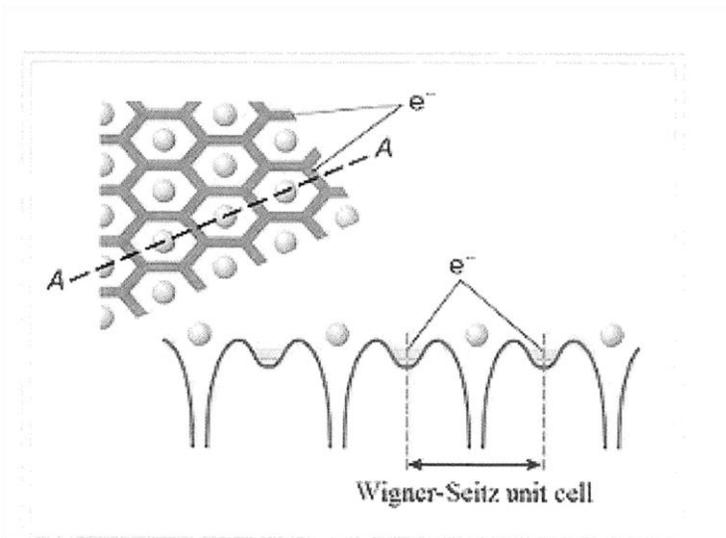

**Fig. 2**. Schematic electron distribution in Rydberg matter. [5]

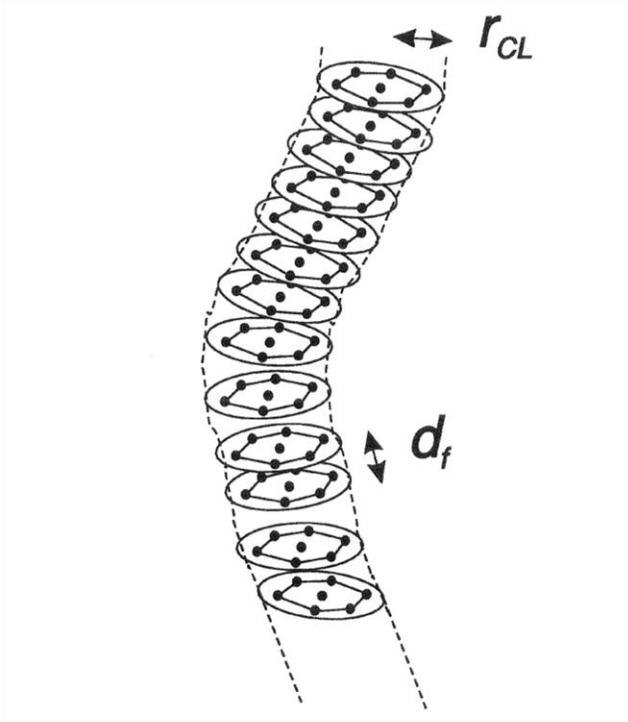

**Fig. 3**. Tower of piled-up Rydberg atoms [6].

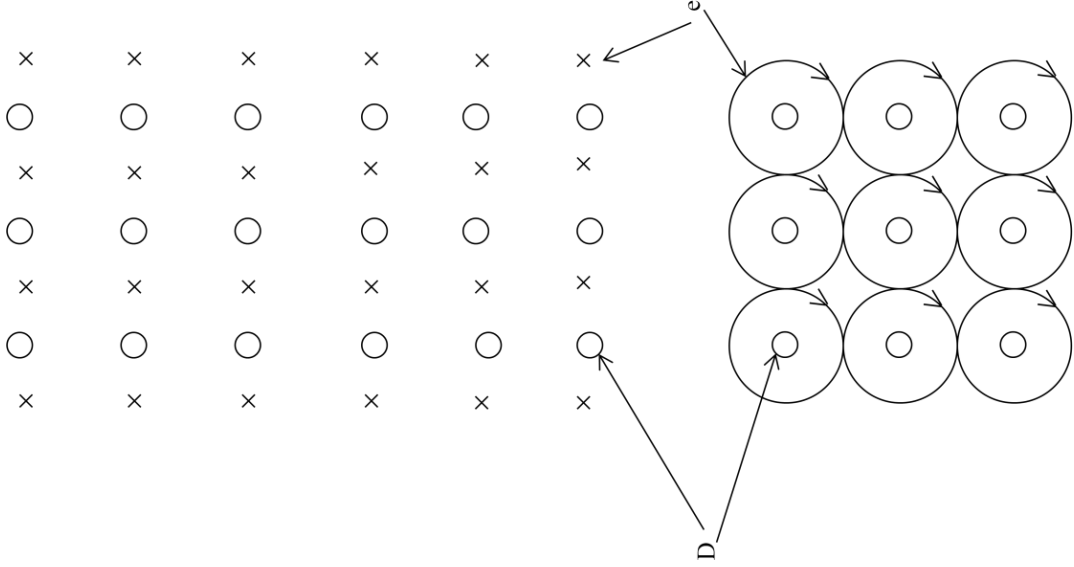

**Fig.4.** Ultradense deuterium- electron vortex tower: D deuterium nuclei, e electron fluid vortices.

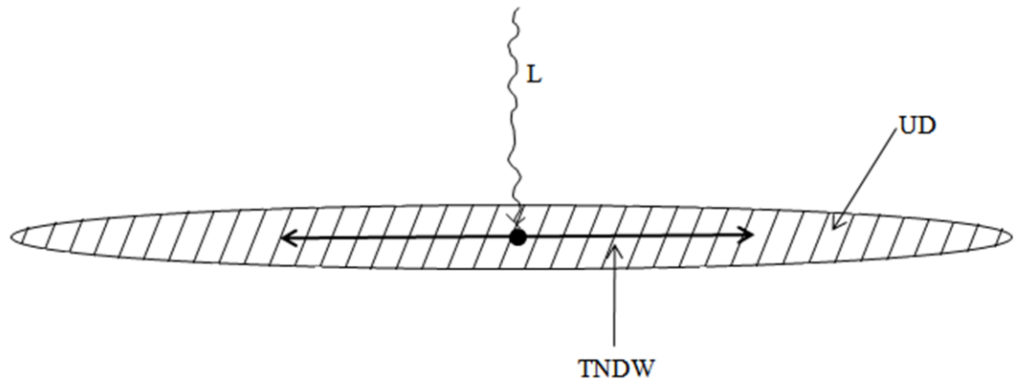

**Fig. 5.** Ultradense Deuterium disc target: UD ultradense deuterium, L petawatt laser or particle beam, TNDW thermonuclear detonation wave.